
\input phyzzx

\REF\aj{
 R. J. Adler and O. C. Jacob,  Submitted to J. Math.  Phys., 1995.}
\refend

\REF\ch{ R. Courant and D. Hilbert, Methods of Mathematical Physics, Vol.2
(Interscience, John Wiley and Sons, New York, 1962); see chapters 5 and 6.}
\refend

\REF\abs{ R. J. Adler, M. Bazin, and M. M. Schiffer, Introduction to General
Relativity, 2nd ed.  (McGraw Hill, New York, 1975); see chapter 4 for the
characteristics of the Maxwell equations and chapter 8 for those of the vacuum
Einstein equations.}
\refend

\REF\ngc{ O. C. Jacob, Phys. Rev. {\bf D50}, 5289 (1994).}
\refend

\REF\pertre{ O. C. Jacob, Phys. Rev. {\bf D51}, 3017 (1995).}
\refend

\REF\mp{ C. Morosi and L. Pizzocchero, J. Math. Phys. {\bf 35}, 2397 (1994).}
\refend

\REF\sovoliev{ V. O. Sovoliev, J. Math. Phys. {\bf 34}, 5747 (1994).}
\refend

\REF\lv{ J. Losa and J. Vives, J. Math. Phys. {\bf 35}, 2856 (1994).}
\refend

\REF\diracqm{ See for example P. A. M. Dirac, Quantum Mechanics, 4th ed.
(Oxford University Press, Clarendon, 1958); chapter 5 contains the classic
discussion of time evolution of quantum states. }
\refend

\REF\bd{ See for example J. D. Bjorken and S. D. Drell, Relativitstic Quantum
Mechanics (McGraw Hill, New York, 1964); chapter 1 introduces the Dirac
equation, chapter 2 discusses Lorentz covariance, and chapter 3 discusses the
free particle solutions. }
\refend

\REF\nsgr{ These null coordinates have been used often and to great advantage
in general relativity when radiation or black hole horizons are concerned. See
for example chapters 4 and 8 of N. D. Birrell and P. C. W. Davies, Quantum
Fields in Curved Space (Cambridge University Press, New York, 1982). }

\REF\bmccpp{ S. J. Brodsky, G. McCartor, H. C. Pauli, and S. S. Pinsky,
Particle World {\bf 3}, no. 3, 109 (1993)}
\refend

\REF\penrose{ R. Penrose, Gen. Rel. and Grav. {\bf 12}, no. 3, 225 (1980).}
\refend

\REF\zhs{ H. M. Zum Hagen and H.-J. Seifert, Gen. Rel. and Grav. {\bf 8}, no.
4, 259 (1977).}
\refend

\REF\kerr{ R. P. Kerr, Phys. Rev. Letters, {\bf 11}, 237 (1963); the Kerr
solution representing  a rotating black hole is built from a metric containing
a null vector; see also ref. 2, chapter 7. }
\refend

\REF\hawking{ S. Hawking, Commun. Math. Phys. {\bf 43}, 199 (1975).}
\refend

\REF\iz{ C. Itzykson and J.-B. Zuber, Quantum Field Theory (McGraw Hill, New
York, 1980); see chap. 1.}
\refend

\PHYSREV
\overfullrule0pt

 \newtoks\slashfraction
 \slashfraction={.13}
 \def\slash#1{\setbox0\hbox{$ #1 $}
 \setbox0\hbox to \the\slashfraction\wd0{\hss \box0}/\box0 }

  \def\Buildrel#1\under#2{\mathrel{\mathop{#2}\limits_{#1}}}

\def\lozenge{\boxit{\hbox to 1.5pt{%
             \vrule height 1pt width 0pt \hfill}}}

 \doublespace
 \pubnum{69xx \cr
 hep-th/9503nnn}
\date{March    1995}
 \pubtype{T}
 \titlepage
 \title{Null Surfaces, Initial Values and
 Evolution Operators for Spinor Fields
 \doeack}
 \author{
 Ovid C. Jacob
\foot{Permanent Address:
\it{
 Department of Physics and Astronomy,
 Sonoma State University,
 Rohnert Park, CA 94928}
}
\foot{jacob@unixhub.slac.stanford.edu
 }
and
Ronald J. Adler\foot{ Permanent Address:
\it{
 Department of Physics and Astronomy,
 San Francisco State University,
 San Francisco, CA 94132}
}
}
 \SLAC

\abstract

We analyze the initial value problem for spinor fields obeying the Dirac
equation, with particular attention to the characteristic surfaces. The
standard Cauchy initial value problem for first order differential equations is
to construct a solution function in a neighborhood of space and time from the
values of the function on a selected initial value surface. On the
characteristic surfaces the solution function may be discontinuous, so the
standard Cauchy construction breaks down. For the Dirac equation the
characteristic surfaces are null surfaces. An alternative version of the
initial value problem may be formulated using null surfaces; the initial value
data needed differs from that of the standard Cauchy problem, and in the case
we here discuss the values of separate components of the spinor function on an
intersecting pair of null surfaces comprise the necessary initial value data.
We present an expression for the construction of a solution from null surface
data; two !
 analogues of the quantum mechanica
l Hamiltonian operator determine the evolution of the system.

\vfill
\centerline{PACS: 11.10 Gh, 11.10Ef, 11.30Er, 11.15-Bt}
\centerline{Submitted to Jour. Math. Phys.}

\chapter{ Introduction}

In reference [\aj] we discussed the Cauchy problem, initial values, and
characteristics of the Klein Gordon equation for scalar fields. In this work we
extend the discussion to include spinor fields obeying the free particle Dirac
equation. Although the approach is the same and frequent references are made to
reference 1 we have attempted to make this paper self-contained.

The standard Cauchy problem for a first order partial differential equation is
to use the values of the solution function on an initial value surface to
determine the values of the function in a neighborhood of the surface [\ch].
Solution of the problem proceeds by showing how the initial value data and the
differential equation determine all the time derivatives of the function on the
initial surface and thus allow a power series development of the function for
future times.

If the solution function has a discontinuity across some special surface then
the standard Cauchy problem cannot be solved using that surface for the initial
value data. Such surfaces are called characteristic surfaces or simply
characteristics.
The characteristics of the Dirac equation for free particles in flat space are
easily shown to be null surfaces. Thus the discontinuities of the solution
functions propagate at the velocity of light, independently of the mass
parameter in the equations. The characteristics of Maxwell's equations, the
Klein Gordon equation, and the vacuum Einstein equations of general relativity
are also all null surfaces [\aj], [\abs].

We first briefly review the standard Cauchy problem for the Dirac equation. We
then obtain the characteristics and solve the analogous problem for the planar
null surfaces. Initial data for this situation consists of the values of
disparate components of the spinor function on the pair of null surfaces. We
give a simple evolution operator expression in terms of null coordinates, which
is a direct analogue of the usual expression of quantum theory, except that two
analogues of the Hamiltonian operator determine the evolution of the solution,
that is are generators of displacements in the null coordinates [\ngc, \pertre,
\mp, \sovoliev, \lv]. The result is remarkably similar to that obtained  for
the scalar field [\aj].

\chapter{Development Of a Spinor Function From a Constant Time Hypersurface}

Since the Dirac equation may be expressed in Hamiltonian form the development
of a solution from initial value data given on a constant time hypersurface is
one of the most familiar problems of physics [\diracqm]. We express the Dirac
equation in Hamiltonian form as [\bd]

$$i\psi_{|t}(\vec{x},t) = [-i\vec{\alpha} . \bigtriangledown +
\beta m]\psi(\vec{x},t) = H \psi(\vec{x},t) \eqn\hdir $$

(We use units in which $\hbar = c = 1$.
The slash notation indicates differentiation with respect to the indicated
variable, in this case time t.) We write a Taylor series expansion as

$$\psi(\vec{x},t) = \psi(\vec{x},t)_{t=0} + t \psi_{|t}(\vec{x},t)_{t=0} +
{t^2\over2!}\psi_{|t|t}(\vec{x},t)_{t=0}+...\eqn\taylord  $$

All of the time derivatives at $t = 0$ may be readily obtained from the Dirac
equation and the value of the spinor function at $t = 0$, which we call
$h(\vec{x})$.
When substituted in the series these give the familiar result

$$\psi(\vec{x},t) = \psi(\vec{x},t)_{t=0} + (-iH t) \psi(\vec{x},t)_{t=0} +
{(-i H t)^2\over2!}\psi(\vec{x},t)_{t=0}+...
 = e^{-iHt}h(\vec{x}) \eqn\taylordd $$

The indicated exponential of the Hamiltonian operator is thereby seen as the
finite time displacement operator.

\chapter{Characteristics of the Dirac Equation}

The characteristics of the Dirac equation are found in the same way as for the
Klein Gordon equation in [\aj], except that since the Dirac equation is first
order in time the problem is even easier. We suppose that the initial value of
the solution is given on a hypersurface S, $t=T(\vec{x})$, by a four component
spinor (see fig. 1)

$$\psi(\vec{x},T(\vec{x})) = h(\vec{x}) \eqn\chard$$

and that the solution is given by a Taylor series expansion

$$\psi(\vec{x},t) = \psi(\vec{x},t)_{t=T} + [t-T(\vec{x})]
\psi_{|t}(\vec{x},t)_{t=T} +
{[t-T(\vec{x})]^2\over2!}\psi_{|t|t}(\vec{x},t)_{t=T}+...\eqn\taylorc  $$

To find the first time derivative in this we first express the gradient of the
initial value h in terms of $t=T(\vec{x})$, as

$$\bigtriangledown h(\vec{x}) = \bigtriangledown \psi(\vec{x},t)_{t=T} +
\psi_{|t}(\vec{x},t)_{t=T}\bigtriangledown T(\vec{x}) \eqn\hsol$$

{}From this and the Dirac equation \hdir  we may write an equation for the time
derivative of $\psi$ on S,

$$i[I - \vec{\alpha} .\bigtriangledown T(\vec{x})]\psi_{|t}(\vec{x},t)_{t=T} -
-i\vec{\alpha}.\bigtriangledown h(\vec{x}) + \beta m
h(\vec{x}) \eqn\psitsol $$

This may be solved for the time derivative if the matrix in square brackets has
an inverse. The determinant of that matrix is easily found to be

$$|I-\vec{\alpha} . \nabla T(\vec{x})| = (1 - \nabla T(\vec{x})^2)^2
\eqn\dirdet $$

Thus the condition that S be a characteristic is that the above quantity be
zero, or

$$\nabla T(\vec{x})^2 = 1 \eqn\chareq $$

This is the same as the characteristic equation for the Klein Gordon equation;
thus the characteristics are null surfaces, and are independent of the mass
parameter in the Dirac equation. The particular null surfaces that we will
study in the remainder of this paper are

$$u = t-x = 0 \quad  	u\quad characteristic\quad or\quad null\quad surface $$
$$v = t+x = 0 \quad 	v\quad characteristic\quad or\quad null\quad surface
 \eqn\chardefs $$

These are shown in figure 2.

\chapter{Construction of a Solution Function in Terms of Null Coordinates}

The covariant form of the Dirac equation is [\bd]

$$\gamma^\alpha i{\partial\over\partial x^{\alpha}} \psi = m \psi
\eqn\direqcov$$

We will study this using the null coordinates (see fig. 2)
$u = t - x$ and $v = t + x$, and suppress dependence on y and z in all that
follows [\nsgr]. Then the gamma matrices are $\gamma^u = \gamma^0 - \gamma^1$
and
$\gamma^v = \gamma^0 + \gamma^1$. The Dirac equation and the gamma matrix
algebra in terms of the null coordinates are

$$[\gamma^u i {\partial\over\partial u} +\gamma^v i {\partial\over\partial
v}]\psi(u,v) = m \psi(u,v) \eqn\dens $$

$${\gamma^u\gamma^v\over4}+{\gamma^v\gamma^u\over4} = 1, \quad (\gamma^u)^2
=(\gamma^v)^2 = 0 \eqn\nsdalg $$

We define a pair of projection operators as

$$\Lambda_u = {\gamma^u\gamma^v\over4} \quad \Lambda_v =
{\gamma^v\gamma^u\over4} \eqn\nsproj $$

{}From \nsdalg the usual projection operator properties follow, that is

$$\Lambda_u + \Lambda_v = 1, \quad \Lambda_u\Lambda_u = \Lambda_u,\quad
\Lambda_v\Lambda_v = \Lambda_v,\quad \Lambda_u\Lambda_v = \Lambda_v\Lambda_u =
0 \eqn\projalg $$

To analyze the Dirac equation on and near the null surfaces $u = 0$ and $v = 0$
we make an expansion in the mass m, since the solutions of the zero mass Dirac
equation are functions of only u or only v. Thus we write

$$\psi(u,v) = \psi^{(0)}(u,v) + m \psi^{(1)}(u,v) +m^2 \psi^{(2)}(u,v) + ...
\eqn\nspsiexp$$

Substitution of this into the Dirac equation gives the following set of
iterative equations

$$[\gamma^u i {\partial\over\partial u} +\gamma^v i {\partial\over\partial
v}]\psi^{(0)}(u,v) = 0,\quad
[\gamma^u i {\partial\over\partial u} +\gamma^v i {\partial\over\partial
v}]\psi^{(n)}(u,v) = m \psi^{(n-1)}(u,v) \eqn\nspsirec $$
Solution of the zero order equation is easy by inspection,

$$\psi^{(0)}(u,v) = \Lambda_u f(u) +\Lambda_v g(v) \eqn\nspsizeroord $$

Here the functions f and g are any differentiable 4-tuple functions of u and v;
we will call these the generating functions. It is instructive to consider a
representation of the gamma matrices in which the projection operators L are
diagonal; then the $\Lambda_u$  contains 1's at entry (1,1) and (2,2), and zero
everywhere else; similarly, $\Lambda_v$ has 1's at entries (3,3) and (4,4) and
zeros everywhere else.

As for  the zero order solution , they are (in transpose notation):

$$\psi^{(0)\dagger}(u,v) = (f_1(u), f_2(u), g_1(v), g_2(v))
\eqn\nspsizeroordsol $$

Thus only two components of f and two components of g enter the solution.

The first order equation \nspsirec is

$$[\gamma^u i {\partial\over\partial u} +\gamma^v i {\partial\over\partial
v}]\psi^{(1)}(u,v) = \Lambda_u f(u) +\Lambda_v g(v) \eqn\nspsifirstord $$

Only a particular solution to this is needed since the homogeneous solution may
be absorbed into the zero order solution. We break the equation into u and v
parts by premultiplying by $\Gamma_u$  and using \dens to get

$$i{\partial\over\partial v}\Lambda_u \psi^{(1)}(u,v) =
{1\over4}\gamma^u\Lambda_v g(v) \eqn\nspsigeq $$

The right side of this is a function of v only, so the left side must be also.
Thus  we may write a solution by integration.

$$\Lambda_u \psi^{(1)}(u,v) = {-i\over4}\gamma^u\Lambda_v \int_{v_{0}} ^v dv'
g(v') \eqn\nspsigsol $$

Here $v_0$ is an arbitrary parameter. In the same way we may obtain a similar
expression for the v projection of the solution,

$$\Lambda_v \psi^{(1)}(u,v) = {-i\over4}\gamma^v\Lambda_u \int_{u_{0}} ^u du'
f(u') \eqn\nspsifsol $$

The sum of (21) and (22) gives the complete first order solution

$$\psi^{(1)}(u,v) = {-i\over4}[\gamma^u\Lambda_v \int_{v_{0}} ^v dv' g(v') +
    \gamma^v\Lambda_u \int_{u_{0}} ^u du' f(u')] \eqn\nspsisol $$

The second order equation (16) may now be written as

$$i{\partial\over\partial v}\Lambda_u \psi^{(2)}(u,v) =
{-i\over4}[\gamma^u\Lambda_v \int_{v_{0}} ^v dv' g(v') + \gamma^v\Lambda_u
\int_{u_{0}} ^u du' f(u')] \eqn\nspsisecord $$

As before we premultiply by $\gamma^u$ to obtain, with use of \nsproj and
\projalg,

$${\partial\over\partial v}\Lambda_u \psi^{(2)}(u,v) = {-1\over4}\Lambda_u
\int_{u_{0}} ^u du' f(u') \eqn\nspsifsec $$

Since the right side of this is a function of only u we may write a particular
solution by inspection,

$$\Lambda_u \psi^{(2)}(u,v) = {-1\over4}(v-v_0)\Lambda_u \int_{u_{0}} ^u du'
f(u') \eqn\nspsifsecsol $$

In the same way we may obtain the v projection of the solution,

$$\Lambda_v \psi^{(2)}(u,v) = {-1\over4}(u-u_0)\Lambda_v \int_{v_{0}} ^v dv'
g(v') \eqn\nspsigsecsol $$

Summing \nspsifsecsol and \nspsigsecsol we have the complete second order
solution,

$$ \psi^{(2)}(u,v) = {-1\over4}(v-v_0)[\Lambda_u \int_{u_{0}} ^u du' f(u') +
    (u-u_0)\Lambda_v \int_{v_{0}} ^v dv' g(v')] \eqn\nspsisecsol $$

The procedure is now clear, and we may continue to all orders; the even orders
are similar in form, and the odd orders are similar in form. The complete
series solution is

$$\psi(u,v) = \sum_{n=0}^{\infty}{[-{m^2\over4}(v-v_0)\Gamma_u]^n\over
n!}(1-{im\over4}\gamma^v\Gamma_u)\Lambda_u f(u) +
 \sum_{j=0}^{\infty}{[-{m^2\over4}(u-u_0)\Gamma_v]^j\over
j!}(1-{im\over4}\gamma^u\Gamma_v)\Lambda_v g(v)  \eqn\nspsi $$

where the multiple integral operator $\Gamma^n_u$ is defined as

$$\Gamma^n_u f(u) = \int_{u_{0}} ^u \int_{u_{0}} ^{u'} ... \int_{u_{0}}
^{u^{n-1}} du^n f(u^n) \eqn\gammadef $$

The series (29) may be readily summed to give a concise expression for the
solution

$$ \psi(u,v) = e^{-{m^2\over4}(v-v_0)\Gamma_u
}(1-{im\over4}\gamma^v\Gamma_u)\Lambda_u f(u) + e^{-{m^2\over4}(u-u_0)\Gamma_v
}(1-{im\over4}\gamma^u\Gamma_v)\Lambda_v g(v)  \eqn\nspsicmp $$

This gives a solution of the free particle Dirac equation for any pair of
generating functions f and g; the generating functions together have only four
independent components however, so this initial value data contains the same
amount of information as that in the standard Cauchy problem.

\chapter{ Initial Values on Null Surfaces}

The generating functions f and g are arbitrary functions, and are simply
related to the initial values of the solution on the null surfaces as we will
now discuss. In this and the following section we  set $u_0 = v_0 = 0$ without
loss of generality.

To get a relation between the initial values of the function y on the null
surfaces and the generating functions f and g we set $u = 0$ and then  $v = 0$
in
\nspsicmp and easily find

$$ \psi_0(u,0) = (1-{im\over4}\gamma^v\Gamma_u)\Lambda_u f(u) + \Gamma_v
\Lambda_v g(0)  \eqn\nspsiuinit $$

$$ \psi_0(0,v) = (1-{im\over4}\gamma^u\Gamma_v)\Lambda_v g(v) + \Gamma_u
\Lambda_u f(0)  \eqn\nspsivinit $$

We wish to make a convenient choice for the values of $f(0)$ and $g(0)$;
taking
 $u = v = 0$ in \nspsiuinit, \nspsivinit  we see that

$$\psi_0 (0,0) = \Lambda_u f(0) +\Lambda_v g(0) \eqn\nspsizero $$

Accordingly we choose $f(0) = g(0) = \psi_0(0,0)$; then the quantities that
appear in the evolution expression \nspsicmp are

$$ (1-{im\over4}\gamma^v\Gamma_u)\Lambda_u f(u)=\psi_0(0,0) - \Lambda_v
\psi_0(0,0))  \eqn\nspsicmpu $$

$$ (1-{im\over4}\gamma^u\Gamma_v)\Lambda_v g(v)=\psi_0(0,0) - \Lambda_u
\psi_0(0,0))  \eqn\nspsicmpv $$

Thus the spinorial expressions in parentheses turn the arbitrary 4-tuple
functions into the indicated initial data on the null surfaces. These
expressions may be inverted if it is desired to obtain f and g.

$$ \Lambda_u f(u) =(1+{im\over4}\gamma^v\Gamma_u)[\psi_0(0,0) - \Lambda_v
\psi_0(0,0)]  \eqn\nspsiucmpu $$

$$ \Lambda_v g(v) =(1+{im\over4}\gamma^u\Gamma_v)[\psi_0(0,0) - \Lambda_u
\psi_0(0,0)]  \eqn\nspsivcmpv $$

In terms of the initial values of the solution the expression (31) now reads

$$  \psi(u,v) = e^{-{m^2\over4}(v-v_0)\Gamma_u
}(\psi_0(0,0)-\Lambda_v\psi_0(0.0)) + e^{-{m^2\over4}(u-u_0)\Gamma_v
}(\psi_0(0,0)-\Lambda_u \psi_0(0,0))  \eqn\nspsicmpct $$

This is a complete expression for the evolution of the solution from its values
on the pair of null surfaces $u = 0$ and $v = 0$. Note that only the square of
the mass appears in the evolution operators.

The form of the solution in \nspsicmpct is very similar to that obtained in
[\aj] for scalar fields obeying the Klein Gordon equation; the two exponential
evolution operators are identical and only the forms of the initial data
expressions are slightly different since the present one contains spin
information. Thus in terms of the null coordinates the dynamics of scalar and
spinor fields is remarkably similar, much more so than in terms of the usual
Cartesian coordinates.

\chapter{ Plane Waves}

We wish to verify the consistency of \nspsicmpct for plane wave solutions of
the Dirac equation. We will show that if the appropriate initial data functions
are put into \nspsicmpct then the solution function generated is the
appropriate plane wave. We write a plane wave solution in Cartesian and null
coordinates as

$$\psi(u,v) = e^{-iEt +ikx}w(E,k) = e^{-i\lambda u -i\tau v}w(\lambda,\tau)
\eqn\nspsipw $$

Here w is the usual Dirac 4-tuple spin function and the null momenta are given
by

$$\lambda ={E+k\over2}, \tau = {E-k\over2}, \lambda\tau ={E^2 - k^2\over4} =
{m^2\over4} = \mu^2 \eqn\nsconst $$

Thus the appropriate initial value quantities that enter the evolution
expression \nspsicmpct are

$$\psi_0(u,0) -\Lambda_v \psi_0(0,0) = (e^{-i\lambda u} - \Lambda_v
)w(\lambda,\tau) $$

$$\psi_0(0,v) -\Lambda_u\psi_0(0,0) = (e^{-i\tau v} -
\Lambda_{u})w(\lambda,\tau) \eqn \nspsipwinit $$

We substitute these expressions into \nspsicmpct, abbreviating for convenience
$B = -i\lambda, C = -i\tau, - {m^2\over4} = BC$, and find

$$\psi(u,v) =
(e^{BCv\Gamma_u}(e^{Bu}-\Lambda_v)w(\lambda,\tau)+(e^{BCu\Gamma_v}(e^{Cv}-\Lambda_u)w(\lambda,\tau) $$

$$ =  (e^{BCv\Gamma_u}e^{Bu}+e^{BCu\Gamma_v}e^{Cv})w(\lambda,\tau)-
F_c(uv)w(\lambda,\tau) \eqn\nspsiexpos $$

Here $F_c$ denotes the function related to a constant generator, defined and
evaluated as

$$F_c(uv) = \sum_{n=0}^{\infty} {(-\mu^2 u v)^2\over(n!)^2} =
e^{-{m^2\over4}v\Gamma_{u}} 1 = e^{-{m^2\over4}u\Gamma_{v}} 1  \eqn\nspsifc$$

Notice that this function is symmetric in u and v. The quantity in parentheses
in \nspsiexpos is straight forward to evaluate; we expand the exponentials in
double series as

$$ (e^{BCv\Gamma_u}e^{Bu}+e^{BCu\Gamma_v}e^{Cv}) = \sum_{n=0,
j=0}^{\infty}{(BCv\Gamma_u)^n\over n!}{(Bu)^j\over j!} + \sum_{m=0,
k=0}^{\infty}{(BCu\Gamma_v)^m\over m!}{(Cv)^k\over k!} \eqn\powersersol $$

It is easy to obtain the operation of $\Gamma^n$ on powers of u,

$$\Gamma_u ^n u^j = {u^{j+n}\over (j+1)(j+2) ... (j+n)} \eqn\gammapowers$$

We substitute this into \powersersol and rearrange summation indices to find

$$e^{BCv\Gamma_u}e^{Bu}+e^{BCu\Gamma_v}e^{Cv}) = \sum_{n\le
j}^{\infty}{(Cv)^n\over n!}{(Bu)^j\over j!} + \sum_{m\ge
k}^{\infty}{(Cv)^m\over m!}{(Bu)^k\over k!}  $$

$$  = \sum_{n=0, j=0}^{\infty}{(Cv)^n\over n!}{(Bu)^j\over j!} + \sum_{m=0,
k=0}^{\infty}{(Cv)^m\over m!}{(Bu)^k\over k!}+\sum_{n=0}^{\infty} {(-\mu^2 u
v)^2\over(n!)^2} = e^{B u + C v} + F_c (uv) \eqn\nsfirstterm  $$

We now combine (40) and (44) to get the complete solution

$$\psi(u,v) = e^{Bu +Cv}w(\lambda,\tau) = e^{-i\lambda u -i\tau v}
w(\lambda,\tau)  \eqn\nspsipwsol $$

which is the correct plane wave solution. We have thus checked the consistency
of a known solution with our formalism. Moreover it follows that the evolution
operator equation \nspsicmpct will produce the correct solution in any case
that may be expanded as a superposition of plane waves.

We note that the constant terms in the initial data expression \nspsicmpu,
\nspsicmpv play a very important role in the formalism and cannot be ignored.

\chapter{ Relation to the Hamiltonian Viewpoint}

The evolution operators in \nspsicmpct are close symbolic analogues of the
usual evolution operator in expression \taylordd, as we will discuss in some
detail.

A Taylor series expansion in $t$ of a function may be expressed symbolically as
an exponential as follows

$$\psi(\vec{x},t) = e^{(t-t_{0}){\partial\over\partial t} }\psi(\vec{x}, t_{0})
\equiv \sum_{n=0}^{\infty} {(t-t_0)^n\over n!} \psi_{|nt}(\vec{x},t) _{t=t_{0}
} \eqn\ethamevol $$

That is the quantity in the exponent generates a time shift. In standard
quantum mechanics the time derivative is related to the energy operator by
${\partial\over\partial t} = -i E$. To define the dynamics the energy operator
in terms of t is identified with the Hamiltonian operator H in terms of x. That
is a Schroedinger equation is postulated, with an operator equivalence

$${\partial\over\partial t} = -i H \eqn\nsschro $$

Then for a wave function the Taylor series expression reads

$$\psi(\vec{x},t) = e^{-i(t-t_{0}) H} \psi(\vec{x},t_0) \eqn\nstaylor$$

which is the standard form for the time evolution operator, as in \taylordd.

We can use the same procedure to obtain our result \nspsicmpct - but only
symbolically and up to a normalization and constant terms. We begin with an
expression for a double Taylor series expansion of a function of u and v in
analogy with \ethamevol.

$$\psi(u,v) = {1\over2}[e^{(v-v_{0}){\partial\over\partial v} }\psi(u, v_{0})
  +e^{(u-u_{0}){\partial\over\partial u} }\psi(u_0, v)]\eqn\nsevol $$

In analogy with the Schroedinger equation we define the dynamics with the Klein
Gordon equation, which in terms of u and v is,

$${\partial^2\over\partial u \partial v} \psi(u,v) = -{m^2\over4} \psi(u,v)
\eqn\nsquadde $$

 - which solutions of the Dirac equation must obey. In analogy with \nsschro
this allows us to identify the v derivative operator with the following
operator in terms of u,

$${\partial\over\partial v} = -{m^2\over4} ({\partial\over\partial u})^{-1}
\equiv -{m^2\over4} \Gamma_u \eqn\nsgammaudef $$

Substituting this into \nsevol we obtain \nspsicmpct  - if we do not concern
ourselves about the normalization factor and the constant terms, which the
symbolic derivation does not seem to explain.

 In conclusion our main result \nspsicmpct presents the evolution of a spinor
field in terms of two analogues of the Hamiltonian operator, and the dynamics
of the spinor field is remarkably similar to that of the scalar field.

\chapter{Acknowlegements}

 We wish to thank S. J. Brodsky and the Stanford Linear Accelerator Center
Theory Group for their hospitality.  M. Weinstein was most helpful in critical
discussions of the manuscript.

\endpage
\refout
\end